# Mechanical properties of graphene oxide: the impact of functional groups

Mahdi Tavakol[1], Abbas Montazeri[2,1,*], S. Hamed Aboutalebi[1,3], Reza Asgari[1,4,*]

[1]School of Nano Science, Institute for Research in Fundamental Sciences, IPM, Tehran, Iran

[2]Computational Nanomaterials Lab (CNL), Faculty of Materials Science and Engineering, K.N. Toosi University of Technology, Tehran, Iran

[3]Condensed Matter National Laboratory, Institute for Research in Fundamental Sciences, IPM, Tehran, Iran

[4]School of Physics, Institute for Research in Fundamental Sciences, IPM, Tehran, Iran

**Abstract**

In the current study, mechanical characteristics of graphene oxide (GO) as a promising substitute of graphene are systematically studied through molecular dynamics simulation. For this purpose, several GO samples having different concentrations of epoxide and hydroxyl functional groups are considered. The results reveal that increasing the epoxide coverage causes a noticeable deterioration in the mechanical characteristics of GO systems. This change is correlated with the increase of the formation of ripples in the structure upon increasing the epoxide coverage. Moreover, investigating the bond lengths in the system, it is concluded that the higher epoxide percentage leads to an increase in the length of single and hybrid resonance bonds leading to an overall deterioration of the mechanical properties of GO samples. Additionally, our results demonstrate that high concentration of functional groups can lead to a negative Poisson's ratio. Increasing the amount of hydroxyl groups shows the same declining effect on the Young's modulus. In a graphene system containing both epoxide and hydroxyl groups, it is deduced that a higher percentage of the former can result in a higher residual strain because of the formation of more ripples within the system.

* Corresponding authors.

E-mail addresses: a_montazeri@kntu.ac.ir (A. Montazeri), asgari@theory.ipm.ac.ir (R. Asgari).

1. Introduction

With many perceived applications in diverse areas ranging from highly sophisticated biological applications [1] and environmental remediation [2] to such niche areas including energy storage [3, 4], printing electronics and catalysts, graphene oxide (GO) and its reduced form (rGO) have manifested themselves as a highly sought-after two-dimensional (2D) material of choice [5-8]. However, central to the realization of many promises of these wonderful 2D materials is the ability to control and manipulate their mechanical properties. Mechanical performance of 2D materials, as the key single criterion in seeking high performance durable materials, is the main governing factor that plays the key role in the process of manufacturing [9], integration into devices and ultimately the performance and lifetime of the final device [10]. This calls for an immediate effort into the investigation and identification of the factors affecting the mechanical performance of GO and rGO. The presence of multi-functional oxide groups exhibits a profound impact on the mechanical features of GO. Therefore, understanding these effects can help us engineer the extent and type of oxygen containing groups to tailor the mechanical properties according to applications. This challenge, if resolved, can have a direct impact on the introduction of novel manufacturing processes along with help in broadening the scope of structural design of practical nanodevices to achieve the maximum potential of this family of 2D crystalline materials in desired applications.

As such, many efforts have been directed to explore the mechanical properties of this new class of materials both experimentally and theoretically [11, 12]. However, performing such measurements have proved to be particularly challenging because of the difficulties in the manipulation of free-standing atomic monolayers [13]. Although, study of the mechanical properties of graphene-based derivatives by first principle calculations have been performed extensively [14, 15], recent experiments have suggested a huge deviation in results, mainly due to the existence of rippling and the effects of temperature. As such, it is strongly suggested to implement statistical mechanics approach along with molecular dynamics (MD) simulations to have a better prediction of the elastic behavior of 2D materials [10]. Motivated by experiments pointing to a remarkable stability of GO with large strains, we explore the effect of the concentration of functional groups on the mechanical properties of GO by carrying out MD simulations. We performed our simulations on different GO samples ranging from fully oxidized to less oxidized samples with different concentrations of epoxide and hydroxyl functional groups, based on the guidelines provided by [16, 17]. Our results demonstrate that the mechanical characteristics of GO systems change noticeably with increasing the concentration of functional groups. Besides, it illustrates that high concentration of functional groups can lead to a negative value of the Poisson's ratio in low strain values. Our results show that in the system having a constant number of surface coverage, increasing the epoxide percentage causes an increase in both the Young's modulus and the residual strain values. Characteristic change in the Carbon-Carbon bonds, an increase in their length and different number of ripples are among the underlying physical reasoning for the difference between pristine graphene and GO. It should be pointed out that while the wrinkled morphology is usually not present in the strained graphene [18, 19], appearance of ripples within the GO surface could provide the physical interpretation of its mechanical properties.

The paper is organized as follows. In section 2, the methodology and the MD approach are discussed. The mechanical properties of the system based on the concentration of functional groups, namely Young's module, residual strain, bond length and Poisson's ratio as functions of epoxide and hydroxyl percentages are presented in section 3. We conclude and summarize our main results in section 4.

## 2. Methodology

In the MD approach, which is a theoretical method for exploring the dynamics of atoms of a crystal interacting for a fixed period of time, the atomic movements and the dynamic evolution of the system are simulated through numerically solving the Newton's second law of motion [20]. In order to deploy the second law, there is a need to obtain the forces exerted on each atom. Accordingly, interatomic potentials are employed to calculate the interatomic forces [21]. Therefore, the first step in the current study was to determine the suitable interatomic potential. For pure graphene sample, considered here as the reference model, the AIREBO potential with the modified cut-off value was implemented [22]. Also, there are several potential functions that can be utilized for studying mechanical properties of GO. They include the second generation of REBO potential calibrated for C-H-O systems [23], CHARMM potential [24], COMB3 [5] and ReaxFF [26] force fields. Fonseca et al. examined the first three potentials for calculating the physical and chemical properties of GO [27]. While the COMB3 potential showed better overall results, the accuracy of the REBO potential in calculating mechanical properties of single layer GO system was better. This finding alongside with the smaller computational cost of the REBO potential persuaded the authors to choose this potential for the current study. The potential file used in these simulations was taken from [28]. As the next step, there was a need for an initial model. According to our best knowledge, there is no software for carrying out the task. Therefore, a C++ program was developed to alternate the structure of pure graphene to that of GO by placing epoxide and hydroxyl groups on its surface. Looping through all the atoms, the specified functional groups were placed onto each atom with a specified chance on the condition that each carbon atom has just four covalent bonds. Hence, the initial model of the GO system was ready for the uniaxial loading simulation. The size of the simulation box was equal to 90*50*10 nm$^3$ composed of ~150000 to ~210000 atoms based on the percentage of functional groups on the GO surface. Simulations were carried out in an NVT ensemble at the temperature of 300 K using 0.1 fs time-step. The LAMMPS software was utilized in all simulations [29].

Since the initial configuration of the constructed samples may not be in the equilibrium state, the system was relaxed for 5000 steps (i.e., 500 ps), before loading. After relaxation, in order to exert the quasi-static uniaxial loading conditions on the sample, the system was incrementally stretched in several steps. During the test, one side of the sample was kept fixed while the other one was stretched by applying an incremental strain of 0.2%. The total size of the fixed and stretched regions was 5 % of the sample length. Additionally, free boundary conditions (BC's) were considered in the axial and lateral directions. It is noted that each loading step consists of two stages. At the first step which lasts for $T_1$ ps, the system is stretched, and then the system is allowed to relax for $T_2$ ps. In this study, different values of $T_1$ and $T_2$ were examined to evaluate which one can imitate the quasi-static loading conditions in a better way. It was found that the accuracy of the results is dominantly dependent on the loading time ($T_1$). In the case of short loading times, longer values of $T_2$ would not produce a reasonable result. To obtain the desired values for these parameters, different schemes were analyzed based on trial and error. Accordingly, the values of 1.6 ps and 0.8 ps were chosen for $T_1$ and $T_2$, respectively, to obtain the best results. Employing these values, final snapshot of the 9% epoxide contained GO system under uniaxial tension is depicted in Fig. 1-a. Having obtained the suitable strain rate, we proceeded to examine the mechanical response of GO nanosheets in the presence of different functional groups. It is worth mentioning that, as discussed in the literature, mechanical features of nanoscale samples are affected by the applied strain rate. However, it was demonstrated that increasing the strain rate would lead to the sudden fracture of the samples near the boundary regions. In contrast, while reducing the strain rate would immensely increase the simulation time, it did not significantly affect the mechanical response of the epoxide covered systems. Accordingly, it was deduced that the introduced values for $T_1$ and $T_2$ were the optimum time-scales to perform our simulations.

### 3. Results and Discussion

GO exhibits considerable variation in properties depending on the degree of oxidation and synthesis methods [30]. It can be functionalized in many ways, depending on the desired application [31, 32]. However, the precise atomic and electronic structure of GO remains largely unknown despite many attempts to unlock the structure of GO [33, 34]. For the sake of simplicity, GO can be considered as a system consisting of monolayer graphene decorated with oxygen functional groups on both basal plane and edges [35]. The common structure of GO consists of hydroxyl and epoxide groups on the basal plane [36].

#### 3.1. Mechanical properties of GO samples with different epoxide coverages

As the first step, GO samples having different epoxide coverages were subjected to uniaxial tensile loading deformation using a series of MD simulations to make use of microcanonical ensemble averages. The corresponding stress-strain curves are depicted in Fig. 1-b. For the sake of comparison, the values representing Young's modulus and residual strain of the samples are summarized in Fig. 1-c. It can be deduced from the figures that increasing the epoxide coverage causes a noticeable decrease in the Young's modulus. The decrease in the Young modulus is ascribed to the change in the character of the carbon-carbon bonds in the system. In pure graphene, all bonds exhibit a hybrid resonance behavior. As this hybrid resonance structure is composed of delocalized π orbitals and σ bonds, it shows higher stiffness compared to single σ bonds. The formation of epoxide between two neighboring carbon atoms eliminates the possibility to have delocalized π orbitals resulting in the formation of single bonds. Thus, some of the bonds lose their stiffness leading to a decrease in the Young's modulus.

It should be noted that the superior mechanical properties of pure graphene are the direct result of the planarity of most bonds. Meanwhile, as seen in Fig. 1-d, 1-e and 1-f, the presence of the epoxide functional groups would lead to the formation of some ripples on the surface of GO sheets leading to a decrease in the Young's modulus due to the reduction of the number of planar bonds. We also performed the unloading simulation for the 9% and 40% epoxide containing GO systems to analyze the plastic behavior of the samples. The loading and unloading routes were the same indicating no sign of any permanent deformation.

Furthermore, the residual strain increased with an increase in the percentage of epoxide groups (Fig. 1-c). This parameter was defined as the size of the initial *healing* region of the stress-strain curves. Interestingly, after the removal of the residual strain, the external loading increases. This is due to the formation of ripples in GO systems, which is more pronounced in the structures having more epoxide coverage. While most of the ripples are present, the GO system cannot withstand external loads causing a negligible stress with increasing the external strain. Accordingly, a higher number of ripples causes an increase in the residual strain as they decrease the planarity of the bonds. Thus, higher epoxide coverage results in higher values of residual strain due to an enhancement in the number of ripples in the GO. Fitting a line into the stress-strain curve of the 40% epoxide covered GO at the strains smaller than 5%, the Young's modulus was found to be about 403 GPa. This value is in good agreement with the experimentally measured values available in the literature [37], which validates the developed tensile code in the current study. Additionally, it is deduced from the figure that there is a residual strain on the system. For strains smaller than 3%, the stress does not change with increasing the strain. It is worth mentioning that our numerical simulations show that there are several ripples available within the sample that will be *quantitatively* discussed later on.

It should be noted that the presence of the epoxide functional groups not only influences the mechanical characteristics of GO samples, but also, more importantly, it affects the fracture event within the discussed 2D nanostructures. For all the epoxide functionalized GO systems, the crack propagation occurred at such a fast pace that the simulation would crash at the onset of the crack formation. After the crash, the MD simulation was restarted considering fixed BCs. In these cases, after introduction of the initial crack, the BCs were kept fixed until the complete breakage. Consequently, it is concluded that the addition of the epoxide functional groups affects the fracture behavior of the graphene sheet sample.

### 3.2. Bond lengths of the epoxide covered GOs

To find out the mechanism governing the deteriorating effects of the epoxide groups on the mechanical characteristics of GO samples, we investigated the variation in the bond lengths of the epoxide covered systems. It should be noted that in pure graphene, there are single and double C-C covalent bonds having a hybrid resonance behavior. The average length of C-C bonds within the structure is equal to 1.42 Å. As there are four atoms covalently connected to some carbon atoms in the GO structure, some of the C-C bonds lose their resonance behavior resembling the characteristics of single bonds. This might be the reason for the deterioration of mechanical properties of graphene after the oxidation process. This phenomenon can also be ascribed to the inability of C-O bonds in bearing the external loads. To explore the matter, the bond lengths of the samples were analyzed at strains of 0 and 25 % using the OVITO software [38].

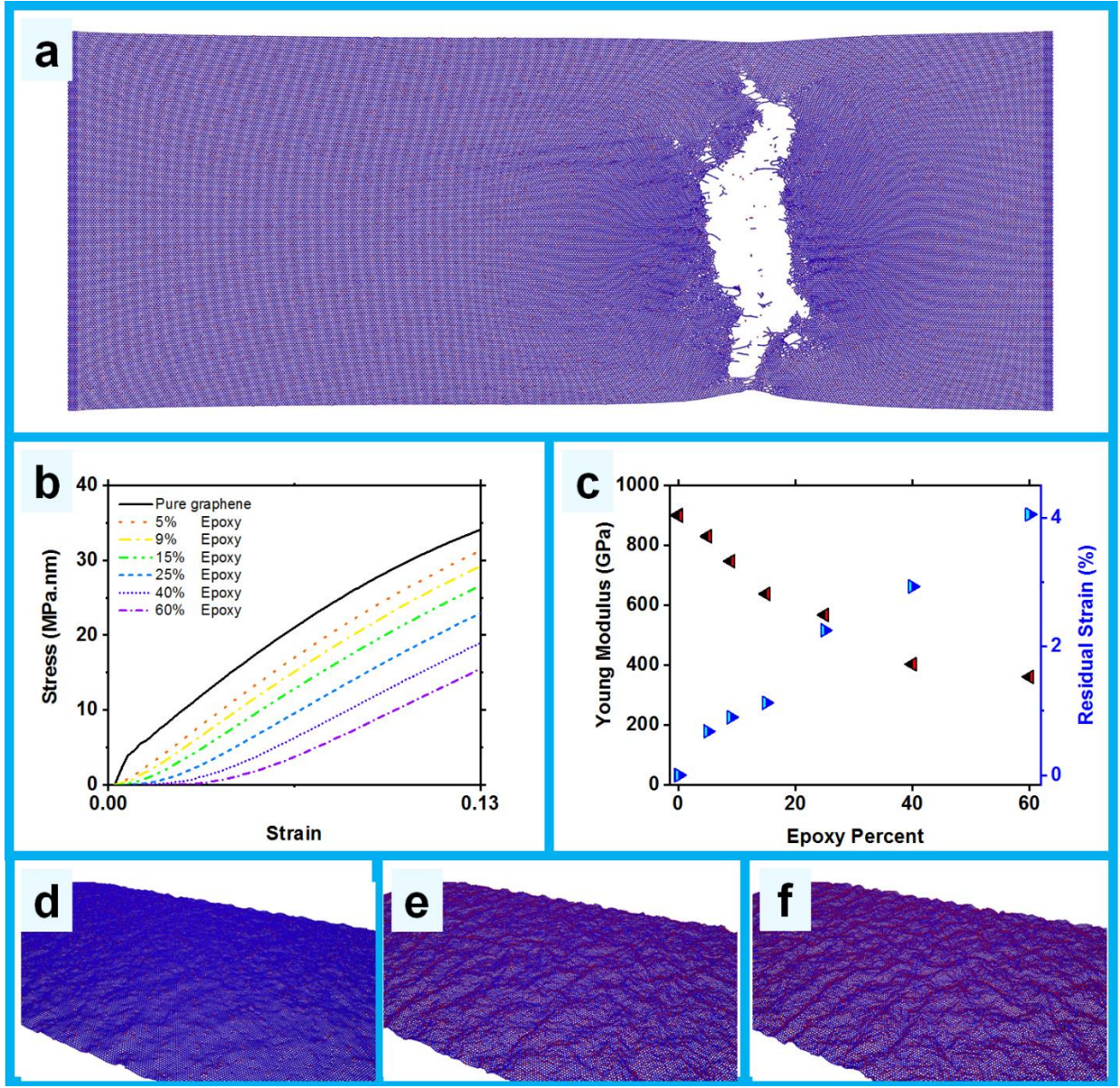

Figure 1 – The mechanical properties of GO based on the percentage of its epoxide functional groups. (a) Fracture of 9% epoxide covered GO system shows that the loading rate and the boundary conditions are appropriate to prevent unphysical fractures at the edges. (b) The stress – strain curves of GO samples having different surface coverages of epoxide functional groups. (c) Variations of Young's modulus and residual strain versus the epoxide percent show corroboration between these two parameters. The equilibrated GO sheets having (d) 5%, (e) 40% and (f) 60% showing an increase in the number of ripples which is the main reason for an enhancement in the residual strain with increasing the epoxide coverage in GO.

The distribution and average value of the bond lengths for different GO systems at the strain of 0 (relaxed structure) are summarized in Fig. 2. The average bond length of the pure graphene sample is equal to 1.43 Å, which is in good agreement with the value of 1.42 Å reported in the literature [39]. Considering the curves in Fig. 2-a, we conclude that there are two types of bonds. The first type having shorter lengths corresponds to the normal C-C bonds in graphene, while the longer bond which is absent in pure graphene, is related to the carbon atoms connected to an epoxide group (Fig. 2-b). As depicted in this figure, increasing the number of the epoxide groups leads to an increase in the number of single bonds. Simultaneously, the bond length of both the hybrid resonance and single bonds is increased resulting in their stiffness reduction (Fig 2-b). These phenomena would manifest themselves in the converse relation between the number of the epoxide groups and the Young's Modulus of GO samples (See Fig. 1-c).

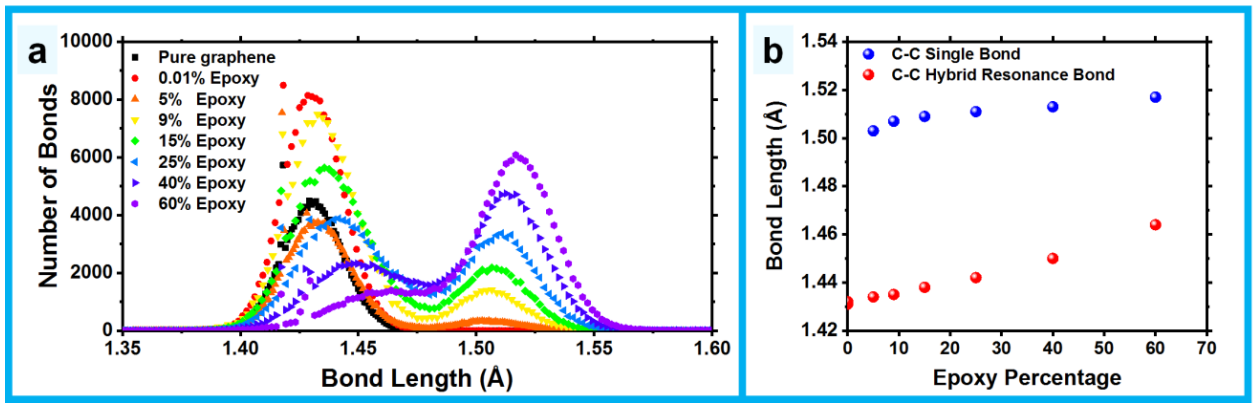

Figure 2. The GO bond length implies a mechanism for the observed changes in the mechanical properties of GO system through the bond length change. (a) C-C bond length distribution for different epoxide coverages of GO systems shows the presence of two different covalent bonds in the system. (b) The changes in the average length of two different bond types with increasing the epoxide percentage showing an increase in the bond length, which leads to a reduction in the bond stiffness.

Additionally, as depicted in Fig. 2-a and 2-b, passing a certain coverage value, all of the bonds change to single type and also, the bond length would no longer be affected by the number of the epoxide groups. Crossing the critical value of 40% epoxide coverage, the mentioned two factors do not have any extra weakening effects on the mechanical properties of GO samples. This is in accordance with the results shown in Fig. 1-c demonstrating that at a determined amount of the epoxide coverage, Young's modulus of GO systems does not change significantly. In a similar way, the C-C bond length distribution corresponding to the strain of 25 % for different GO systems is shown in Fig SI-1 (see the supplementary). Contrary to the previous case, at larger strain values, the distinction between different kinds of the bonds in a GO system is eliminated. This parameter then falls within the range of single C-C bonds. Thus, at large strains, hybrid resonance bonds break into single bonds. Consequently, graphene and GO systems have only single C-C bonds. Considering the smaller stiffness of the π bond of the double bond of $sp^2$ hybridized carbon atoms, this conclusion makes sense. To be more precise, each double bond consists of a σ bond, which is the same as single bonds and a π bond. Owing to some geometrical constraints, the latter has smaller stiffness [40]. Therefore, at larger strains, the π bond breaks and as a result, there are just σ bonds available within the GO system, representing single bonds.

Moreover, the bond length distribution of C-O covalent bonds at the strains of 0 % and 25 % are shown in Fig SI-2-a and SI-2-b (see the supplementary). It is found that neither the epoxide percentage nor the strain changes the bond length. As a result, the epoxide group does not carry any external forces in uniaxial loading of GO. The negligible values of the stress on the oxygen atoms in the stress contours illustrated in Fig SI-2-c further verify this issue. As such, it can be concluded that higher epoxide percentages cause an increase in the C-C bond length, change of hybrid bonds to single one and the elimination of the resonance phenomenon consequently leading to the deterioration of mechanical properties.

### 3.3. Mechanical properties of the hydroxyl covered GOs

In order to model hydroxyl functionalized GO systems, the United Atom (UA) model was utilized [41]. In this model, the hydrogen atoms are considered as a part of their covalently connected carbon atoms. The mass of hydrogen is added to the oxygen to take into account its dynamics. However, there is a possibility that the oxygen atom makes covalent bonds with the neighboring carbon atoms forming an epoxide group. To prevent this from happening, the interactions between oxygen atoms of the hydroxyl group with other carbon atoms except for those they are connected to were excluded. Subsequently, the hydroxyl groups were modeled in the current study as oxygen atoms with larger mass having reduced reactivity.

The stress-strain curves for different hydroxyl covered GO systems are illustrated in Fig 3-a. Interestingly, increasing the hydroxyl groups coverage results in a decrease in the Young's modulus with very small residual strain penalty. The equilibrated 40 % hydroxyl covered GO system is shown in Fig 3-b. As illustrated in this figure, the number of ripples is smaller in the hydroxyl covered GO systems with respect to their epoxide covered counterparts. This might be the reason for the lack of the residual strains in GO systems containing hydroxyl groups. Moreover, the Young's modulus value of 452.5 GPa calculated for 40 % hydroxyl covered GO system is in agreement with the experimentally measured values [37].

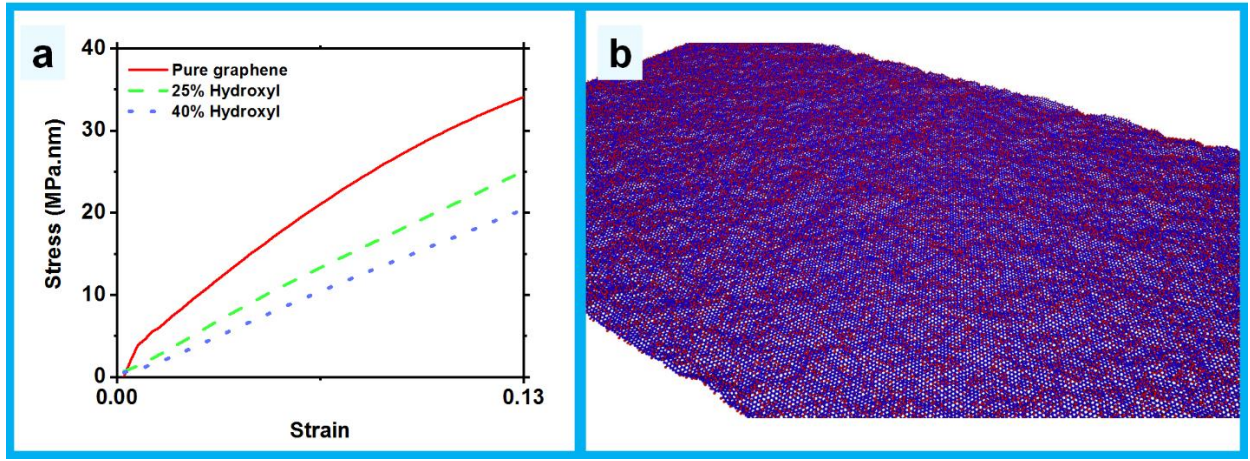

Figure 3. Mechanical properties of the hydroxyl covered GO systems. (a) Stress – strain curves for different percentages of hydroxyl functional groups illustrate a deterioration of the mechanical properties of the system with increasing the hydroxyl coverage along with the absence of residual strain. (b) GO system covered with 40% of the hydroxyl groups. Lower amount of surface ripples is held accountable for the lack of the residual strain in the hydroxyl covered systems.

### 3.4. Mechanical properties of both hydroxyl and epoxide containing GO systems

To study the cooperative effect of both hydroxyl and epoxide functional groups on the mechanical properties of GO samples, the systems containing both groups were simulated. To this end, the uniaxial loading tests were simulated for five different case studies namely; 40% epoxide, 30% epoxide – 10% hydroxyl, 20% epoxide – 20% hydroxyl, 10% epoxide – 30% hydroxyl, and 40% hydroxyl groups. The corresponding stress-strain curves are given in Fig. 4-a. To facilitate better comparisons, the mechanical properties are also summarized in Fig 4-b. As seen, upon increasing the epoxide percentage, there is a decrease in the value of elastic modulus. As such, at a fixed amount of surface functional groups, systems having a higher percentage of hydroxyls show better elastic properties. It is noted that the value of the Young's modulus of the GO system containing 40% epoxide groups is calculated as 137 GPa.nm which is in good agreement with the experimentally measured value of 145.32 ± 16.38 GPa.nm [37]. However, the residual strain shows an increase as a result of increasing the epoxide percentage. As mentioned before, this increase is attributed to the higher amount of ripples for the epoxide covered GOs. Comparing the equilibrium configuration of GO system containing 5%, 40% and 60% of epoxy as illustrated in Figs 1-d, 1-e, and 1-f with the GO system of 40% hydroxyl represented in Fig 3-b, corroborates the idea of increasing the ripples with an increase of the epoxide percentage.

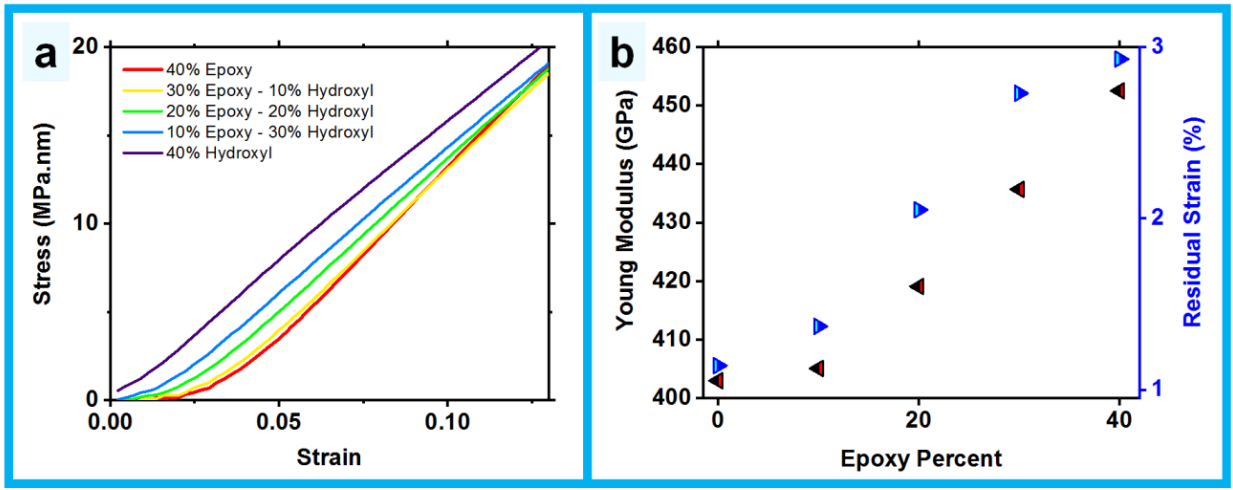

Figure 4. Mechanical properties of the systems containing 40% surface coverage of the epoxide and hydroxyl functional groups. (a) Stress-strain curves for different composition of the epoxide and hydroxyl groups. (b) Young's modulus and residual strain versus the epoxide coverage. Higher epoxide percentage (i.e. lower hydroxyl coverage) causes an increase in both Young's modulus and residual strain value.

### 3.5. Poisson's Ratio

Poisson's ratio, which is the negative of the ratio of transverse strain to axial strain, was measured:

$$\gamma = -\frac{d\vartheta_y}{d\vartheta_x} \tag{1}$$

where $\gamma$, $\vartheta_x$ and $\vartheta_y$ represent the Poisson's ratio, the axial and transverse strains.

Accordingly, the widths of the samples were also obtained to calculate the lateral strain. The simulation snapshot of the GO system having 60% epoxide at the strain of 0.2% is depicted in Fig. 5-a. This figure shows an increase in the sample width implying a negative Poisson's ratio. Unfortunately, since the lateral strain was very small, there was huge noise intensified with the differentiation process. Therefore, a heavy noise reduction process was performed on the output values. The Poisson's ratio for the epoxide covered GO systems is depicted in Fig. 5-b. Note that the overall trend for the pure graphene is the same as the one reported in [42]. However, the results obtained here are larger than the values calculated in that reference. This might be due to several reasons. In the mentioned study, periodic boundary conditions (PBCs) have been utilized in the lateral direction. Thus, the width of the sample is very large and consequently, the determined Poisson's ratio is smaller than its real value. Meanwhile, in the present work, simple BCs were implemented in the lateral direction to mimic the samples in a more realistic way. Furthermore, geometrical characteristics of the simulation cells were not the same in these two works. The size of the sample was $10*10$ nm$^2$ in [42], while in the current study, a sample size of $90*50$ nm$^2$ was simulated.

As depicted in Fig. 5-b, for the strain values belonging to the small deformation regime, there is a decrease in the value of the Poisson's ratio with increasing the epoxide coverage. Consequently, the Poisson's ratio for systems having higher than 25% epoxide surface functional groups is negative. The decreasing trend of the Poisson's ratio can be attributed to the existence of residual strain. As Fig. 1-c shows, the value of residual strain enhances with increasing the epoxide percentage. Higher residual strain values lend the system a tendency for having larger deformations in order to release that strain. As a result, when the system is under specific strain rate in the longitudinal direction, it is free to move in the cross sectional direction. So, the deformation in the lateral directions can be higher than the longitudinal one in the systems having higher epoxide coverage due to their higher residual strain.

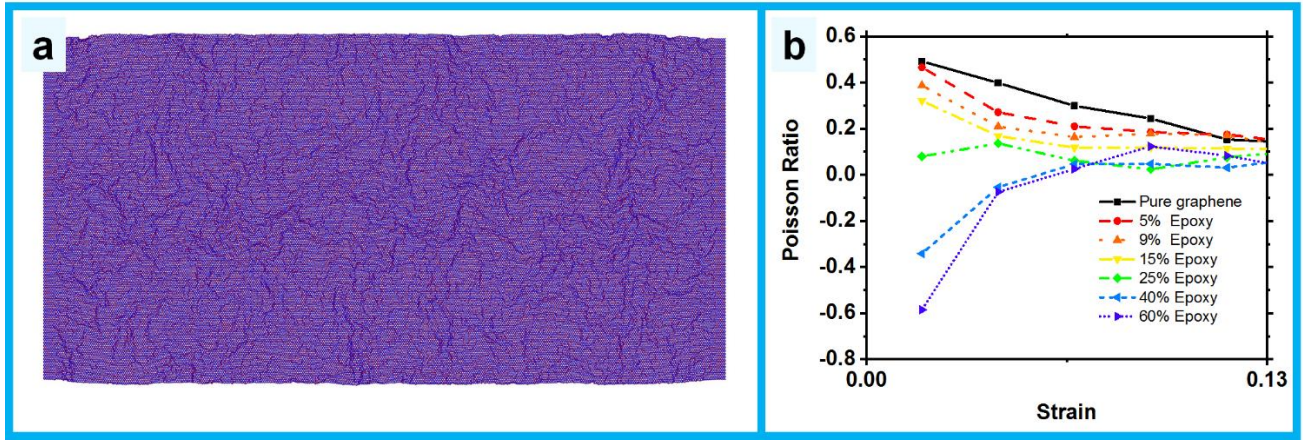

Figure 5. The lateral elongation of the epoxide covered GO system during uniaxial tension. (a) Snapshot of the 60% epoxide covered system at the strain of 0.2%. (b) The Poisson's ratio versus strain for different epoxide covered GO systems.

### 4. Conclusion

In the current study, the mechanical properties of graphene Oxide (GO) were studied using MD simulations. First, several strategies in simulating the uniaxial tension test were tested to obtain the proper one to avoid the unphysical simulation results. It was deduced that large relaxation time in the uniaxial loading simulation cannot compensate for the small loading time. Actually, the simulation results were more sensitive to the latter than the former and there is a need for large loading time. This was followed by the inspection of the effects of the epoxide coverage on the mechanical behavior of GO samples. The simulation results illustrated that the mechanical properties deteriorate with increasing the epoxide percentage. Also, there is a residual strain within GO samples that increases with increasing the number of epoxide groups. An enhancement in the number of ripples was considered as a reason for the observed ascending trend. In addition, to shed light on the effect of the epoxide groups on GO, the bond lengths of C-C and C-O bonds were thoroughly investigated at the strain of 0 and 25 %. It was concluded that in the relaxed state, there were two types of single and hybrid resonance bonds. Higher epoxide coverage led to an increase in the number of single bonds and also an increase in the length of both types of bonds. These two reasons were enumerated as the main factors influencing the deterioration of GO mechanical properties in the presence of epoxide groups. Moreover, at the strain of 25 %, there were no hybrid resonance bonds. Afterwards, the mechanical properties of the hydroxyl covered GO systems alongside with systems having both of the functional groups were explored. For the pure hydroxyl contained GO systems, all the mechanical properties weakened upon increasing the number of functional groups. Besides, there were negligible residual strains owing to their smaller number of ripples within the relaxed structure. For the systems having both functional groups, higher percentage of epoxide in the fixed number of total functional groups caused smaller Young's modulus and larger residual strain. Finally, Poisson's ratio of the epoxide covered systems was inspected. It was deduced that higher residual strains owing to the higher number of epoxide groups can cause negative Poisson's ratio. This work provides a universal guideline to tune the mechanical properties of oxide based graphene derivatives as a function of the level of oxidation and the introduced functional groups. Moreover, with changing the percentage of functional groups, the mechanical properties of GO systems can be tailored for specific applications based on the results obtained in the current research. To complete the discussion, taking into account the crack initiation and growth in the studied systems would be worthwhile. Hence, as an ongoing work, providing an in-depth analysis of the crack propagation schemes in the introduced covered GO samples seems like an interesting idea.

## 5. Acknowledgments

This work is supported by the Iran Science Elites Federation. The authors would like to thank X. Mu and T. Luo from the University of Notre Dame for sharing the CO.airbo potential file.

**References**

1. L. Zhang, J. Xia, Q. Zhao, L. Liu, Z. Zhang, *Functional graphene oxide as a nanocarrier for controlled loading and targeted delivery of mixed anticancer drugs*. Small, 2010. 6(4): p. 537-544.

2. R. K. Joshi, P. Carbone, F. C. Wang, V. G. Kravets, Y. Su, I. V. Grigorieva, et al., *Precise and ultrafast molecular sieving through graphene oxide membranes*. Science, 2014. 343(6172): p. 752-754.

3. S. H. Aboutalebi, R. Jalili, D. Esrafilzadeh, M. Salari, Z. Gholamvand, S. A. Yamini, et al., *High-performance multifunctional graphene yarns: toward wearable all-carbon energy storage textiles*. ACS nano, 2014. **8**(3): p. 2456-2466.

4. S. H. Aboutalebi, S. Aminorroaya-Yamini, I. Nevirkovets, K. Konstantinov, H. K. Liu, *Enhanced hydrogen storage in graphene oxide-MWCNTs composite at room temperature*. Advanced Energy Materials, 2012. 2(12): p. 1439-1446.

5. D. Voiry, J. Yang, J. Kupferberg, R. Fullon, C. Lee, H.Y. Jeong, et al., *High-quality graphene via microwave reduction of solution-exfoliated graphene oxide*. Science, 2016. 353(6306): p. 1413-1416.

6. S. Pei, Q. Wei, K. Huang, H. M. Cheng, W. Ren, *Green synthesis of graphene oxide by seconds timescale water electrolytic oxidation*. Nature Communications, 2018. 9(1): p. 145.

7. B. Liu, K. Zhou, *Recent progress on graphene-analogous 2D nanomaterials: properties, modeling and applications*. Progress in Materials Science, 2019. **100**: p. 99-169.

8. R. Jalili, D. Esrafilzadeh, S. H. Aboutalebi, Y. M. Sabri, A. E. Kandjani, S. K. Bhargava, et al., *Silicon as a ubiquitous contaminant in graphene derivatives with significant impact on device performance.* Nature Communications, 2018. **9**(1): p. 5070.

9. S. Naficy, R. Jalili, S. H. Aboutalebi, R. A. Gorkin, K. Konstantinov, P. C. Innis, et al., *Graphene oxide dispersions: tuning rheology to enable fabrication.* Materials Horizons, 2014. **1**(3): p. 326-331.

10. D. Akinwande, C. J. Brennan, J. S. Bunch, P. Egberts, J. R.Felts, H. Gao, et al., *A review on mechanics and mechanical properties of 2D materials—Graphene and beyond.* Extreme Mechanics Letters, 2017. **13**: p. 42-77.

11. H. Zhan, D. Guo, G. Xie, *Two-dimensional layered materials: from mechanical and coupling properties towards applications in electronics.* Nanoscale, 2019. **11**(28): p. 13181-13212.

12. C. Androulidakis, K. Zhang, M. Robertson, S. Tawfick, *Tailoring the mechanical properties of 2D materials and heterostructures.* 2D Materials, 2018. **5**(3): p. 032005.

13. G. Cao, H. Gao, *Mechanical properties characterization of two-dimensional materials via nanoindentation experiments.* Progress in Materials Science, 2019.

14. J. E. Barrios Vargasm, B. Mortazavi, A. W. Cummings, R. Martinez-Gordillo, M. Pruneda, L. Colombo, et al., *Electrical and thermal transport in coplanar polycrystalline graphene-hBN heterostructures.* Nano Letters, 2017. 17(3): p. 1660-1664.

15. B. Mortazavi, M. Shahrokhi, M. Raeisi, X. Zhuang, L. F. C. Pereira, T. Rabczuk, *Outstanding strength, optical characteristics and thermal conductivity of graphene-like BC3 and BC6N semiconductors.* Carbon, 2019. 149: p. 733-742.

16. M. Nováček, O. Jankovský, J. Luxa, D. Sedmidubský, M. Pumera, V. Fila, et al., *Tuning of graphene oxide composition by multiple oxidations for carbon dioxide storage and capture of toxic metals.* Journal of Materials Chemistry A, 2017. **5**(6): p. 2739-2748.

17. A. M. Dimiev, J.M. Tour, *Mechanism of graphene oxide formation.* ACS nano, 2014. **8**(3): p. 3060-3068.

18. N. Abedpour, R. Asgari, F. Guinea, *Strains and pseudomagnetic fields in circular graphene rings.* Physical Review B, 2011. **84**(11): p. 115437.

19. M. A. H. Vozmediano, M. I. Katsnelson, F. Guinea, *Gauge fields in graphene.* Physics Reports, 2010. **496**(4-5): p. 109-148.

20. M. Tavakol, M. Mahnama, R. Naghdabadi, *Shock wave sintering of Al/SiC metal matrix nano-composites: A molecular dynamics study.* Computational Materials Science, 2016. **125**: p. 255-262.

21. D.C. Rapaport, *The Art of Molecular Dynamics Simulation*. 2004: Cambridge University Press.

22. B. Mortazavi, Z. Fan, L. F. C. Pereira, A. Harju, T. Rabczuk, *Amorphized graphene: A stiff material with low thermal conductivity.* Carbon, 2016. 103: p. 318-326

23. B. Ni, K. -H. Lee, S. B. Sinnott, *A reactive empirical bond order (REBO) potential for hydrocarbon–oxygen interactions.* Journal of Physics: Condensed Matter, 2004. **16**(41): p. 7261-7275.

24. J. Huang, A.D. Jr. MacKerell, *CHARMM36 all-atom additive protein force field: Validation based on comparison to NMR data.* Journal of Computational Chemistry, 2013. **34**(25): p. 2135-2145.

25. T. Liang, T. –R. Shan, Y. –T. Cheng, B. D. Devine, M. Noordhoek, Y. Li, et al., *Classical atomistic simulations of surfaces and heterogeneous interfaces with the charge-optimized many body (COMB) potentials.* Materials Science and Engineering: R: Reports, 2013. **74**(9): p. 255-279.

26. Chenoweth, K., A.C.T. van Duin, and W.A. Goddard, *ReaxFF Reactive Force Field for Molecular Dynamics Simulations of Hydrocarbon Oxidation.* The Journal of Physical Chemistry A, 2008. **112**(5): p. 1040-1053.

27. A. F. Fonseca, T. Liang, D. Zhang, K. Choudhary, S. B. Sinnott, *Probing the accuracy of reactive and non-reactive force fields to describe physical and chemical properties of graphene-oxide.* Computational Materials Science, 2016. **114**: p. 236-243.

28. X. Mu, X. Wu, T. Zhang, D. B. Go, T. Luo, *Thermal Transport in Graphene Oxide – From Ballistic Extreme to Amorphous Limit*. Scientific Reports, 2014. 4: p. 3909.

29. S. Plimpton, *Fast Parallel Algorithms for Short-Range Molecular Dynamics.* Journal of Computational Physics, 1995. **117**(1): p. 1-19.

30. S. H. Aboutalebi, A. T. Chidembo, M. Salari, K. Konstantinov, D. Wexler, H. K. Liu, S. X. Dou, *Comparison of GO, GO/MWCNTs composite and MWCNTs as potential electrode materials for supercapacitors.* Energy & Environmental Science, 2011. **4**(5): p. 1855-1865.

31. M. Chakraborty, M. S. J. Hashmi, *Graphene as a Material–An Overview of Its Properties and Characteristics and Development Potential for Practical Applications.* 2018. 10.1016/B978-0-12-803581-8.10319-

32. R. Jalili, S. H. Aboutalebi, D. Esrafilzadeh, R. L. Shepherd, J. Chen, S. Aminorroaya-Yamini, et al., *Scalable one-step wet-spinning of graphene fibers and yarns from liquid crystalline dispersions of graphene oxide: towards multifunctional textiles*. Advanced Functional Materials, 2013. 23(43): p. 5345-5354.

33. D. Pacilé, J. C. Meyer, A. Fraile Rodríguez, M. Papagno, C. Gómez-Navarro, R. S. Sundaram, et al., *Electronic properties and atomic structure of graphene oxide membranes.* Carbon, 2011. **49**(3): p. 966-972.

34. S. Saxena, T. A. Tyson, S. Shukla, E. Negusse, H. Chen, J. Bai, *Investigation of structural and electronic properties of graphene oxide.* Applied Physics Letters, 2011. **99**(1): p. 013104.

35. K. A. Mkhoyan, W. Contryman, J. Silcox, D. A. Stewart, G. Eda, C. Mattevi, et al., *Atomic and electronic structure of graphene-oxide*. Nano Letters, 2009. 9(3): p. 1058-1063.

36. G. Eda, M. Chhowalla, *Chemically derived graphene oxide: towards large‐area thin‐film electronics and optoelectronics.* Advanced Materials, 2010. **22**(22): p. 2392-2415.

37. J. W. Suk, R. D. Piner, J. An, R. S. Ruoff, *Mechanical properties of monolayer graphene oxide*. ACS nano, 2010. 4(11): p. 6557-6564.

38. A. Stukowski, *Visualization and analysis of atomistic simulation data with OVITO–the Open Visualization Tool.* Modelling and Simulation in Materials Science and Engineering, 2009. **18**(1): p. 015012.

39. O. V. Yazyev, L. Helm, *Defect-induced magnetism in graphene.* Physical Review B, 2007. **75**(12): p. 125408.

40. Streitwieser, A., et al., *Introduction to organic chemistry*. 1992: Macmillan New York.

41. D. Y. Yoon, G.D. Smith, T. Matsuda, *A comparison of a united atom and an explicit atom model in simulations of polymethylene.* The Journal of Chemical Physics, 1993. **98**(12): p. 10037-10043.

42. J. –W. Jiang, T. Chang, X. Guo, H. S. Park, *Intrinsic negative Poisson's ratio for single-layer graphene*. Nano Letters, 2016. 16(8): p. 5286-5290.